\documentclass[]{spie}
\usepackage{times}
\usepackage{natbib}
\usepackage{graphicx}

\long\def\Ignore#1{\relax}

\title{ Simulations of ELT-GMCAO performance for deep field observations}
\author
       {E. Portaluri,$^{1,2}$\thanks{E-mail:elisa.portaluri@inaf.it}
    V. Viotto,$^{1,2}$
    R. Ragazzoni,$^{1,2}$
    C. Arcidiacono,$^{1,2}$
    M. Bergomi,$^{1,2}$\and
    M. Dima,$^{1,2}$
    D. Greggio,$^{1,2}$
   J. Farinato,$^{1,2}$ and
   D. Magrin$^{1,2}$\\
$^{1}$INAF--Osservatorio Astronomico di Padova, vicolo dell'Osservatorio 5, I-35122 Padova, Italy
\\$^{2}$ADONI - Laboratorio Nazionale Ottiche Adattive - Italy}

\begin{document}

\date{Accepted xxx Received xxx ; in original form \today}

\maketitle

\label{firstpage}
\begin{abstract}  
The Global-Multi Conjugated Adaptive Optics (GMCAO) approach offers an alternative way to correct an adequate scientific Field of View (FoV) using only natural guide stars (NGSs) to extremely large ground-based telescopes. Thus, even in the absence of laser guide stars, a GMCAO-equipped ELT-like telescope can achieve optimal performance in terms of Strehl Ratio (SR), retrieving impressive results in studying star-poor fields, as in the cases of the deep field observations. 
The benefits and usability of GMCAO have been demonstrated by studying 6000 mock high redshift galaxies in the {\it Chandra Deep Field South} region. However, a systematic study simulating observations in several portions of the sky is mandatory to have a robust statistic of the GMCAO performance. Technical, tomographic and astrophysical parameters, discussed here, are given as inputs to GIUSTO, an IDL-based code that estimates the SR over the considered field, and the results are analyzed with statistical considerations.
 The best performance is obtained using stars that are relatively close to the Scientific FoV; therefore, the SR correlates with the mean off-axis position of NGSs, as expected, while their magnitude plays a secondary role. This study concludes that the SRs correlate linearly with the galactic latitude, as also expected.

 Because of the lack of natural guide stars needed for low-order aberration sensing, the GMCAO confirms as a promising technique to observe regions that can not be studied without the use of laser beacons. It represents a robust alternative way or a risk mitigation strategy for laser approaches on the ELTs.
  \end{abstract}


\section{Introduction}
\label{sec:intro}

 One of the main questions that the next generation of telescopes will have to answer is how galaxies form and evolve in a cosmological context, likely using deep field observations. The need of an extraordinary resolution to advance in this research field can be satisfied either by space telescopes or by the next generation of extremely large telescopes (ELTs, i.e., the European Extremely Large Telescope, \citealt{Gilmozzi2007}, the Giant Magellan Telescope, \citealt{Johns2008}, and the Thirty Meter Telescope, \citealt{Szeto2008}), which will play a key role because of their angular resolution and their capability in collecting the light of faint sources.
Especially in the latter case, a high signal-to-noise ratio is needed, and it can not be obtained efficiently with post-processing or Fourier extraction techniques, like speckle or lucky imaging. Still it is reachable only with dedicated Adaptive Optics (AO) modules, where diffraction-limited performances are  possible.

One way to exploit ELTs potentiality is using the Global Multi-Conjugated Adaptive Optics (GMCAO) concept \citep{Ragazzoni2010},  a natural guide stars (NGSs)-based technique that we proposed to ESO as a risk mitigation strategy for the European ELT and that was encouraged after a feasibility study \citep{Bergomi2014}.
In fact, laser guide stars (LGS)-based systems \citep{Foy1985,Rigaut2014} have the advantage that they can correct the FoV uniformly, but even if they are becoming a more reliable solution nowadays (due to the reduction of cost and risk in the recent years), their intrinsic characteristics still imply some limitations. Among them, we only mention the cone effect \citep{Foy2000}; the tilt anisoplanatism \citep{Rigaut1992}, as a consequence of the non-sensitivity to low-order modes and the spot elongation and relative truncation \citep{Muller2011} when sensed by a Shack-Hartmann LGS wavefront sensor. All these issues translate into error budget terms and some other well-known associated criticalities \citep{Fried1995,Pfrommer2009,Diolaiti2012}.

GMCAO corrects  the optical turbulence for a Scientific FoV (e.g., 2 arcmin) taking advantage of a very wide technical FoV (e.g. 10 arcmin, the field used to search for NGSs) to look for AO-suitable NGSs, and thus increasing the sky coverage. This FoV extension is possible only for the next generation of ELTs because of the wide collecting-area, which ensures a larger superposition of the footprints up to high altitudes \citep{Viotto2014}, and is the key difference between GMCAO and well-known MCAO- or LTAO-equipped systems \citep{Beckers1988,Beckers1989}.  

In its concept, GMCAO foresees the use of numerical entities, called Virtual Deformable Mirrors (VDMs) to retrieve information about the atmosphere and correct the images from the  turbulence, driving the real Deformable Mirrors (DMs). Moreover, to increase the wavefront sensor (WFS) sensitivity (to look for fainter references), GMCAO relies on the Very Linear version of the pyramid WFS \citep{Ragazzoni2010}. The latter works in a quasi-open loop mode, saving sensitivity and linearity thanks to the introduction of an internal compensation loop,  with a DM installed in front of the pyramid on a non-common path channel of the optical design, in order to bring the pyramid to its best working regime.

To predict GMCAO performance in terms of Strehl Ratio (SR), we have built an end-to-end numerical simulator, {\it GMCAO Interactive Data Language Unreleased Simulation TOol}, (GIUSTO). It that starts computing the atmosphere on the collected wavefronts and ends with an estimation of the SR in the observed region of the sky \citep[in preparation, hereafter V20]{Viotto2020}. It combines several technical and astrophysical user-defined inputs applying the GMCAO technique, which represents an alternative (or, as already said, a risk mitigation) to the more “classical” MCAO approach. We recall here that the SR parameter is linked to the image quality,  representing a measure of how good is the correction. It can be measured as the peak of the Point Spread Function (PSF) from a point source in the aberrated image intensity over the peak of the ideal one, i.e., using an ideal optical system limited only by diffraction.

We have already shown how a GMCAO-equipped ELT can carry out deep observations successfully (\citealt{Portaluri2015}, \citealt[hereafter EP17]{Portaluri2017}) by simulating a $K$-band mock image of $500\times500$~arcsec$^2$ with 6000 high redshift galaxies (of different sizes and masses, 3 h of integration time) in the {\it Chandra Deep Field South} region. The SR distribution over the area was quite uniform, with a median of $<{\rm SR}>=0.127\pm0.050$  (at 2.2 $\mu$m ) and gave us the chance to detect 99.7\% of the early-type galaxies and 89.4\% of the late-type galaxies with SExtractor \citep{Bertin1996}. Analyzing the outcome of that work, we can consider ${\rm SR}=0.08$ as the lower limit to define successful detection. Below that value, we could not identify late-type galaxies with $M\le10^9$~M$_{\odot}$ at a redshift $z\ge2.75$. In that paper, we also performed the photometric decomposition of the surface brightness profile of each galaxy with Galfit \citep{Peng2002} to distinguish their morphology and calculate their size.

 To get a more robust statistics of the GMCAO performance, here we are extending the previous work by analyzing 22 regions in the sky, each $500\times500$~arcsec$^2$ wide, corresponding to 22 well-known deep-field surveys. These particular areas are of great interest for extragalactic science because of the very low bright stars density. Moreover, they represent a way to understand the limitations of the GMCAO technique and its feasibility, without resorting to the use of LGS.

After presenting the input parameters and constraints that we assumed for running the GIUSTO code (Section~\ref{sec:input}), we describe the sample selection (Section~\ref{sec:overview}), and performance estimations both on geometrical and purely AO perspective for an ELT-like telescope (Section~\ref{sec:performance}).  Section~\ref{sec:conclusion} presents the results and conclusions.

\section{System and GMCAO parameters for the simulations}
\label{sec:input}
As the aim of this work is to increase the statistical significance of EP17,  showing the feasibility of astrophysical investigations with a GMCAO-equipped ELT-like telescope, we considered a number of actual star-poor regions, already studied in well-known deep-field surveys. There, we computed the available number and position of NGSs, and by consequences, the SR achievable  after the AO correction. 

For this purpose, we used an updated version of the Interactive Data Language simulation tool described in V20, that we called GIUSTO, which estimates the performance of a GMCAO system. The code can be applied to a general extremely large telescope. However we tested it mainly with the ELT parameters for different directions in the sky  and provided SRs. 
The output SR depends on several input parameters, such as the choice of the atmospheric behavior (that we assumed to be frozen), the simulated system layout, number, positions, and magnitude of NGSs, technical and scientific FoVs sizes and number and conjugation altitude of VDMs.
Table~\ref{tab:giusto} summarizes the main parameters we adopted, dividing into categories as defined in V20,  where a proper discussion and full tests are presented. In the next subsections, we focus on the accuracy and the consequences of the assumptions we made.

\begin{table}
  \center
\caption{Main parameters used for GIUSTO. Column~1: Parameter. Column~2: Corresponding value. Column~3: Unit of measure.}  
\label{tab:giusto}
\begin{tabular}{lcc} 
\hline
\noalign{\smallskip}
\multicolumn{3}{c}{Reference stars}\\
\noalign{\smallskip}
\hline
Number of NGSs                 & 6               &         \\
Position/Magnitude NGSs        & USNO-B1 catalog &         \\
Limit magnitude                & 18            & mag (R)     \\
\hline
\noalign{\smallskip}
\multicolumn{3}{c}{Tomography}\\
\noalign{\smallskip}
\hline
Screen scale      & 0.1  &[m/pixel]\\
Number of VDMs    & 6    &         \\
\hline
\noalign{\smallskip}
\multicolumn{3}{c}{System}\\
\noalign{\smallskip}
\hline
Sensing wavelength                & $0.5$        & [$\mu$m]\\
Scientific wavelength             & $2.2$        & [$\mu$m]\\
Entrance pupil diameter           & 39           & [m]\\
Probes proximity (min)            & 10           & [arcsec]\\
Technical FoV diameter            & 600          & [arcsec]\\
Central rejected FoV diameter     & 120          & [arcsec]\\
Scientific FoV                    & 120          & [arcsec]\\
Number of fields                  & 100          &\\
Number of DMs                     & 3            & \\
DMs conjugated altitudes          & 0; 4; 12.7   & [km]\\
\hline
\noalign{\smallskip}
\multicolumn{3}{c}{Atmosphere}\\
\noalign{\smallskip}
\hline
Atmospheric profile      & 35   & [layers]\\
Fried Radius             & 0.14 & [m] \\
L$_0$ max                & 25   & [m] \\
l$_0$ min                & 0.001 & [m]\\
Atmospheric spectrum     & von Karman & \\
\hline
\end{tabular}
\end{table} 

In summary, we considered an un-obstructed 39-m circular pupil, a sensing wavelength of 500 nm, a technical FoV of 600 arcsec for searching up to 6 NGSs from the USNO-B1 catalog, with a limit magnitude $R=18$. A circle with a diameter of 120 arcsec in the central area of the technical FoV is not accessible to probes; therefore the algoritm rejects NGSs inside this radius. Moreover, we discarded stars with a separation below 10 arcsec for practical reasons, assuming that a specific mechanical clearance is required to observe close stars with two probes simultaneously. With the nominal $F/17.7$, this mechanical separation translates into about 33 mm. This limitation is somehow conservative because it would be possible to overcome it, for instance, having probes moving on different focal planes, as it has been for the three Shack-Hartmann pick-up probes in MAD \citep{Marchetti2005}.
In a field where stars outnumber the available probes, the code selects the best 6 stars optimizing the correction uniformity.

The adopted AO system design takes advantage of 3 deformable mirrors conjugated to fixed altitudes (0, 4, and 12.7 km), equipped each with 80 actuators along the metapupil diameter, defined by the scientific FoV (2 arcmin).
The 6 GMCAO VDMs are conjugated to the ground and 5 more altitudes, chosen after an optimization that considered the atmospheric profile and the NGSs characteristics (position and magnitudes).
The characteristics of the ELT instrumentation, as currently designed by \citet{Ramsay2018}, drive the selection of the parameters above. However, the work can be generalized to all the next-generation of giant telescope, just changing some input parameters.

Each simulation is composed by 100 sub-fields (that we call ``sectors''), which would be observed sequentially in a possible observing campaign, potentially using different sets of NGSs. They are arranged in a $10\times10$ grid, with a FoV of $50\times50$ arcsec$^2$ each, comparable with the MICADO FoV ($\approx 53\times53$ arcsec$^2$, \citealt{Gilmozzi2007,Davies2016}), and centered on the coordinates listed in the following section.

Each simulation covers an overall field of about $8.3\times8.3$ arcmin$^2$ for a total investigation area of about 1528 arcmin$^2$ (2200 sectors).

It might help if it were described as sub-fields that would be observed sequentially,
> potentially using different sets of NGS.

 \subsection{Sensitivity to turbulence profiles}
\label{sec:atmosphere}

\begin{figure}
  \center
\includegraphics[bb=150 370 580 830,width=9cm]{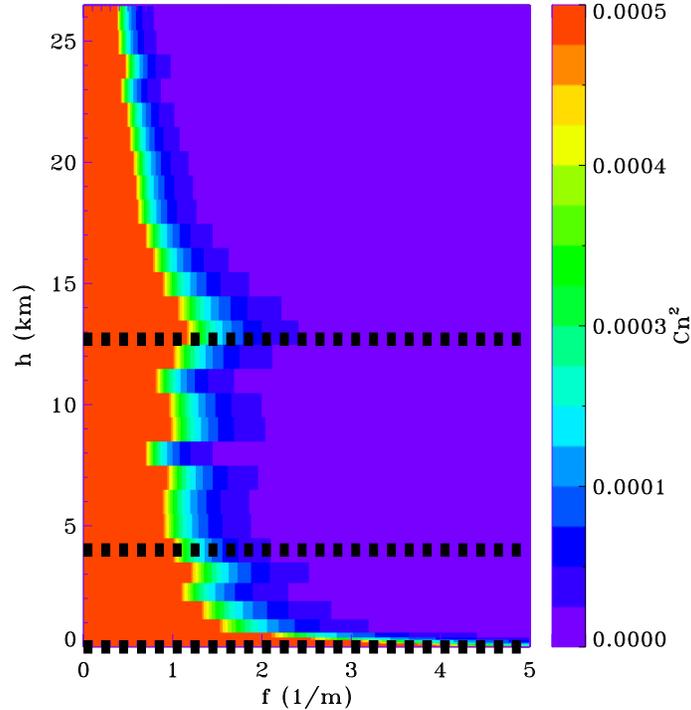}
\caption{$C_n^2$ vertical 3D-profile of the atmosphere considering 35 layers. The horizontal dotted lines indicate the conjugation altitudes of the ELT DMs: 0, 4, and 12.7 km.}
\label{fig:35l}
\end{figure}

The atmospheric profile we considered (Figure~\ref{fig:35l}) is composed of 35 layers, encompassing significant turbulence until an altitude of 26.5 km. We assumed a Fried parameter $r_0=0.129$ m at 500 nm (at $\theta=30^{\circ}$ of zenithal distance, 0.14 m at Zenith). The code builds the turbulence layers as phase screens with power spectrum density following von Karman distribution. The inner and outer scales were $l_0=0.001$ m and $L_0=25$ m, respectively.

The multiple layers approach considers the wavefront error generated by turbulent layers and DM conjugation altitude mismatch, a layer vertical distribution sensitivity test is discussed in \citet{LeLouarn2002}. Here we show that 35 layers are sufficient for the 6 VDM GMCAO performance characterization. 
Between the variety of possibilities, that model has become the reference, being representative of the expected observing conditions at Cerro Armazones and was adopted in several works regarding ELT simulations and predictions \citep{Sarazin2013,Basden2014a,Basden2016,Fusco2010,Costille2012}.
\citet{Sarazin2013} describes both the 40 layers we used in P17 and the 35 layers we adopted here. 
We should also mention that some studies considered a 9-layer profile, as in the cases of \citet{LeLouarn2012} and \citet{Basden2014b}.
However, the effects of a given turbulence profile on the results are still a matter of discussion in the current literature (see \citealt{Helin2015} for an overview).

Moreover, the variation of the profile is not the only parameter playing a role in the game, as the overall seeing modulates the total turbulence power by a significant factor.
For example, \citet{Basden2016} showed that the change of the observed seeing distribution, even using the same profile, would lead to variations larger than an order of magnitude.
Seeing scales singificantly with the airmass ($cos(\theta)^{3/5}$), pushing AO observations close to the meridian passage and selecting targets toward Zenith: an observational and performance trade-off shows that a distance of 30 deg (adopted in this work) may be representative of a typical case.

All these discussions are less relevant for SCAO techniques because, correcting along a single direction, the depth of focus becomes infinite and the different layers are not disentangled. However the corrections get worse as the distance from the reference star increases, because of the vertical turbulence profile is different if we consider the anisoplanatic angle.

Nevertheless, we tested, for a specific case, the dependence of the results with respect to the choice of the model.

\subsection{System efficiency and error budget}

A performance curve is usually employed to quantify the wavefront response of a single WFS, generally in terms of equivalent SR of the residuals as a function of the source magnitude.
In this study, we assumed the performance curve of the {\it First Light AO}, FLAO \citep{Esposito2010a,Esposito2010b}, which was derived from on-sky measurements of existing single conjugated AO systems in closed loop at the Large Binocular Telescope, as a measure of the correction throughput of the system, and properly re-scaled at the adopted wavelength using the Mar\'echal approximation \citep{Marechal1947}.

Actually, in the GMCAO approach, each WFS arm includes a SCAO-like loop, closed between the pyramid and a local DM, to optimize the PSF shape on the pin of the pyramid. So, we may assume that the loop performance of each of these channels is comparable to the FLAO ones, derived from on-sky measurements, provided to consider similar geometrical configuration of the sensor.

Since the available FLAO input performance curve is defined only in terms of WF error rms, it does not contain any information on the residual power spectrum.
Its information is used by the code to introduce a noise pattern, whose amplitude is retrieved from the WFS response curve, onto each WFS measurement, depending on the selected NGS magnitude.
Therefore, during the code development, different noise spectra have been compared and are presented in V20, showing negligible variations of the performance achieved.
 We define the noise model as a combination of an atmospheric power spectrum density, normalized according the star magnitude, and a function mimicking the high frequency part of the turbulence, invisible for the WFS. 

As the GMCAO performance strictly depends on the characteristics of the input SCAO response curve, such an assumption deserves some further comments. The FLAO system employs an actuator density approaching one actuator per $\sim 30$ cm, as projected on the entrance pupil, which would translate onto each SCAO arm for any single WFS with about 140 actuators on a diameter. As this should be accomplished only for an extremely limited FoV, the physical pitch of the actuators can be conveniently rescaled, but nowadays, even MEMS-like DMs technology, are not at such level of overall number of actuators.
However, the results of the GMCAO sky coverage presented in this work rely on the faint-end part of the stellar distribution ($89\%$ of the stars used are in the range between 14 to 18 mag, as discussed in Section~\ref{sec:discussion}), where FLAO system uses only limited number of modes. 
In fact, the sensitivity of these kind of WFSs are dominated by their behaviour around the knees of the curve (around $R\approx 14$ mag, cfr. Figure~2 and Table~1 in \citealt{Esposito2012}).
A thorough inspection of the curve, at various intervals, reveals that the spatial sampling on the pupil, used during the operations for modelling out the WFS behaviour,
is limited by a factor $\approx 3$ with respect to the bright-end performance. This translates into a need for spatial sampling on the local DMs proportionally lowered, about 8 actuators on the diameter.

Furthermore, one should recall that the pyramid closed-loop gain scales with larger telescope diameters (\citealt{Ragazzoni1999}, \citealt{Peter2010}, and for limited SR regimes \citealt{Viotto2016}) as such gain is mostly obtained through low-order modes rather than the high-order ones, although this advantage has not be considered in the scaling laws here.
  Therefore, to some extent, our assumption can be considered a sort of conservative approach to the WFS error estimation.
 
 It is also worth noting that the FLAO response curve used as a basis for the estimation of the wavefront measurement errors was obtained with the former e2V CCD39 detector. Recently, it has been refurbished with a new EMCCD among the upgrades foreseen by the Single conjugated adaptive Optics Upgrade for LBT (SOUL, \citealt{Pinna2016}). In the forthcoming observing periods, LBT will take advantage of the FLAO upgrades: a future extension of our work may exploit the new data.

\section{Deep field selection}
\label{sec:overview}
 The star-poorness is a mandatory condition to select sky regions for deep-field  observations. For this reason, this research topic represents a good test-bench for investigating the performance of the GMCAO technique that takes adavantage of very large FoV for AO reference selection.

This requirement on field selection applies to deep-field observation planning with other considerations, such as low Galactic extinction, declinations range allowing to observe the region over given exposure time, and the efficiency while imaging contiguous viewing zones.
Since this work focuses on simulating the performance of an ELT-GMCAO system for general deep field observations and wants to extend the statistical significance of P17 using a reasonable number of fields, meeting the requirements above goes beyond the aim of our work, while they may be assessed from an observational and operational point of view.
Thefefore, our results are only strictly dependent on the efficiency of the GMCAO correction, with some other assumptions discussed in Section~\ref{sec:input}, and they are not biased by any operative conditions.

   In conclusion, we selected 22 fields as targets of our simulations among those already observed in the framework of deep-field observations and survey campaigns with both space- and ground-based telescopes. We performed this selection using existing literature (\citealt{Ferguson2000,Bowyer2000,Cristiani2001,Brandt2005,Djorgovski2013}). We highlight here that those surveys were studied in several bands for different purposes and are characterized by several magnitude depths. This characteristic matched our aim to have a composite sample of fields that can be used to assess the scientific performance of the GMCAO in the case of deep-field observations.

   Table~\ref{tab:surveys} shows a list of the fields considered for the GMCAO  simulations with the coordinates and names of the corresponding deep-field surveys.

\begin{table*}
  \centering
\caption{List of the  regions of the GMCAO simulations. Column~1: Name of the GMCAO field. Column~2: Deep-field survey acronym corresponding to the GMCAO field. Column~3: Right Ascension and Declination of the central coordinates used by GIUSTO.  Column~4: Galactic longitude and latitude of the central pointing. More information can be found in \citet{Portaluri2018}.}
\label{tab:surveys}
\begin{tabular}{llcc} 
\hline
\noalign{\smallskip}
GMCAO field  & Survey names      &  RA, Dec &  l,b   \\
              &                    & [$^{\circ}$,$^{\circ}$] & [$^{\circ}$,$^{\circ}$]\\
\noalign{\smallskip}
\hline
\noalign{\smallskip}
GMCAO \#1 & CDFS                 &  53.12,-27.81      &    223.57,     -54.44  \\    
          & GEMS                 &                    &    	 \\                     
          & GOODS-S              &                    &    	 \\                     
          & CANDELS: GOODS-S     &                    &    	 \\                     
          & HUDF                 &                    &    	 \\                     
          & HUDF9                &                    &    	 \\                     
          & HUDF12               &                    &    	 \\                     
          & XDF                  &                    &    	 \\                     
GMCAO \#2 & COSMOS               &  150.12, 2.21      &    236.82,      42.12  \\    
          & CANDELS: COSMOS      &                    &                       \\ 	                     
          & CFHTLS-D2            &                    &  	              \\       
GMCAO \#3 & HDF-S                &  338.25, -60.55    &    328.24,     -49.22  \\     
GMCAO \#4 & HDF                  &  189.20, 62.22     &    125.89,      54.83  \\     
          & GOODS-N              &                    &                       \\ 	                     
          & CANDELS: GOODS-N     &                    &   	              \\       
GMCAO \#5 & CANDELS: UDS         &  34.41, -5.20      &    169.89,     -59.96  \\                       
GMCAO \#6 & AEGIS                &  214.25, 52.00     &    96.55,      60.02   \\     
          & EGSS                 &                    &                       \\           
          & GSS                  &                    &                       \\           
          & CFHTLS-D3            &                    &                       \\           
          & CANDELS: EGS         &                    &                       \\            
GMCAO \#7 & CFHTLS-D1            &  36.50, -4.49      &     171.99,     -58.05 \\                                            
GMCAO \#8 & CFHTLS-D4            &  333.88, -17.73    &     39.27,     -52.93  \\    
GMCAO \#9 & SubaruDF             &  201.16, 27.49     &      37.65,      82.62 \\	 
GMCAO \#10 & DLS-1               &  13.35, 12.57      &     123.69,     -50.30 \\    
GMCAO \#11 & DLS-2               &  139.50, 30.00     &     196.17,      43.47 \\    
GMCAO \#12 & DLS-3               &  80.00, -49.00     &     255.51,     -34.82 \\    
GMCAO \#13 & DLS-4               &  163.00, -5.00     &     256.48,      46.82 \\    
GMCAO \#14 & DLS-5               &  208.75, -10.00    &     327.62,      49.80 \\    
GMCAO \#15 & DLS-6               &  32.50, -4.50      &     166.01,     -60.62 \\    
GMCAO \#16 & DLS-7               &  218.00, 34.28     &     57.46,     67.30  \\    
GMCAO \#17 & NDWFS-Bootes        &  217.5, 34.50      &     58.25,      67.66  \\    
GMCAO \#18 & NDWFS-Cetus 	 &  31.87, -4.74      &      165.28,     -61.19\\	     			     
GMCAO \#19 & UKIDSS-DXS1         &  36.25, -4.50      &     171.66,     -58.22 \\    
           & UKIDSS-UDS          &                    &                       \\           
GMCAO \#20 & UKIDSS-DXS2         &  164.25, 57.67     &     148.39,      53.43 \\     
GMCAO \#21 & UKIDSS-DXS3         &  242.50, 54.00     &      83.59,     45.05  \\ 
GMCAO \#22 & UKIDSS-DXS4         &  334.25, 0.33      &      63.02,    -43.85  \\    
\noalign{\smallskip}  
\hline           
\end{tabular}
\end{table*} 


\section{Analysis of the performance of a GMCAO-equipped ELT}
\label{sec:performance}

\subsection{Performance estimations with GIUSTO}
\begin{figure*}
  \center
\includegraphics[bb=47 777 398 1170,width=8cm,height=8.5cm]{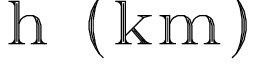}
\caption{Histogram of the VDMs conjugation altitudes distribution, as derived on $10\times10$ fields investigated of the 22 GMCAO simulations. In this plot we excluded the VDM coupled with the ground for graphical reason. The vertical dashed lines indicate the average value for the corresponding altitude of each VDM.}
\label{fig:histohVDM}
\end{figure*}

For each sector of all fields, GIUSTO optimized the altitudes of the VDMs, always keeping one conjugated to the ground (VDM~$\#1$). As expected, the optimal conjugation altitudes of the VDMs vary, depending on the NGSs star geometry. However we can see a common trend, which is linked, of course, to the $C_n^2$, as shown in Figure~\ref{fig:histohVDM}. In particular, for the 35-layers atmosphere, and excluding the VDM~$\#1$, the others are conjugated to average altitudes of 3.7, 9.1, 11.7, 15.9, and 24.4 km respectively.

Their measurements were then projected onto the 3 real DMs, using as optimization criterion the SR performance in the Scientific FoV area.
Table~\ref{tab:SRresults} shows the statistics of the results obtained with this analysis, including the average SR and its standard deviation in all the 100 sectors and the maximum SR value reached. We computes the SR for K band at 2.2 micron. Other values represent:
\begin{itemize}
\item the percentage of the region with ${\rm SR}\ge0.127$ (${\rm SR}_{0.127}$), which is the median value of the SR distribution found by P17; i.e., GMCAO \#1 has ${\rm SR}\ge0.127$ in about 21$\%$ of the area

\item for each field simulation the percentage of the region with ${\rm SR}<0.08$ (${\rm SR}_{0.08}$), which we assumed as a lower limit for the success of the detection (see Section~\ref{sec:intro}); i.e., GMCAO \#1 has ${\rm SR}<0.08$ in about 47$\%$ of the area.
\end{itemize}

Figure~\ref{fig:SR} shows the final map, and the averaged SR with the standard deviation inside each sector for a sub-sample of the simulations.
This sub-sample includes the fields: GMCAO \#4, \#9, \#10, and \#12 for reasons that we discuss in Section~\ref{sec:discussion}. 
The strong discontinuities inside the sectors are due to the dithering technique that we adopted to build such a map, selecting the best value of SR for each sector.
Since noise depends on star brightness, two adjacent fields, partially overlapping, may include diverse constellation with or without bright stars showing in this way different top SR.

We were able to find 6 stars in all the fields except in 7 sectors of the field GMCAO \#9 and 5 sectors of the field GMCAO \#17.

  \begin{table}
  \centering
\caption{SR analysis for the GMCAO simulations campaign. Column~1: Name of the GMCAO field. Column~2 and 3: Average and standard deviation of the SR obtained in all the $10\times10$ sectors of the field. Column~4: Max SR value in the region. Column~5 and 6: Percentage of the field with ${\rm SR}>0.127$ and ${\rm SR}<0.08$. Note that here the values are referred between all the sectors of each field and are different from those reported in Figure~\ref{fig:SR} that show the statistics within one sector} 
\label{tab:SRresults}
\begin{tabular}{lrrrrr} 
\hline
\noalign{\smallskip}
GMCAO field  & $<{\rm SR,tot}>$      &  $\sigma_{\rm SR}$ & max$_{\rm SR}$ & ${\rm SR}_{0.127}$ & ${\rm SR}_{0.08}$   \\
\noalign{\smallskip}
\hline
\noalign{\smallskip}
GMCAO \#1    &   0.09   &   0.05  &    0.31 & 20.9  &   46.9 \\   	   
GMCAO \#2    &   0.19   &   0.05  &    0.32 & 92.9  &   0.5  \\   	   
GMCAO \#3    &   0.22   &   0.05  &    0.34 & 96.5  &   0.3  \\      
GMCAO \#4    &   0.11   &   0.05  &    0.27 & 29.1  &  34.1  \\        
GMCAO \#5    &   0.15   &   0.05  &    0.32 & 62.8  &   7.1  \\       
GMCAO \#6    &   0.17   &   0.06  &    0.36 & 72.6  &   4.5  \\       
GMCAO \#7    &   0.14   &   0.05  &    0.28 & 57.6  &   7.4  \\ 
GMCAO \#8    &   0.15   &   0.05  &    0.29 & 66.4  &   6.8  \\ 
GMCAO \#9    &   0.11   &   0.05  &    0.27 & 28.9  &  30.5  \\ 
GMCAO \#10   &   0.18   &   0.06  &    0.37 & 82.6  &   1.0  \\      
GMCAO \#11   &   0.19   &   0.06  &    0.33 & 84.6  &   1.0  \\      
GMCAO \#12   &   0.25   &   0.05  &    0.38 & 99.8  &   0.0  \\      
GMCAO \#13   &   0.19   &   0.04  &    0.32 & 92.7  &   0.8  \\      
GMCAO \#14   &   0.19   &   0.06  &    0.39 & 80.9  &   6.9  \\      
GMCAO \#15   &   0.19   &   0.05  &    0.33 & 91.9  &   0.5  \\      
GMCAO \#16   &   0.18   &   0.05  &    0.35 & 86.4  &   0.9  \\      
GMCAO \#17   &   0.13   &   0.05  &    0.31 & 51.2  &  23.4  \\
GMCAO \#18   &   0.15   &   0.06  &    0.33 & 61.3  &  16.7  \\
GMCAO \#19   &   0.15   &   0.06  &    0.34 & 60.0  &  16.5  \\
GMCAO \#20   &   0.15   &   0.06  &    0.33 & 60.3  &  12.9  \\
GMCAO \#21   &   0.24   &   0.04  &    0.35 & 99.7  &   0.0  \\ 
GMCAO \#22   &   0.20   &   0.06  &    0.39 & 87.4  &   1.0  \\     
\noalign{\smallskip}  
\hline           
\end{tabular}
\end{table} 

\begin{figure*}
\hbox{\includegraphics[bb=54 355 338 925,width=8.7cm]{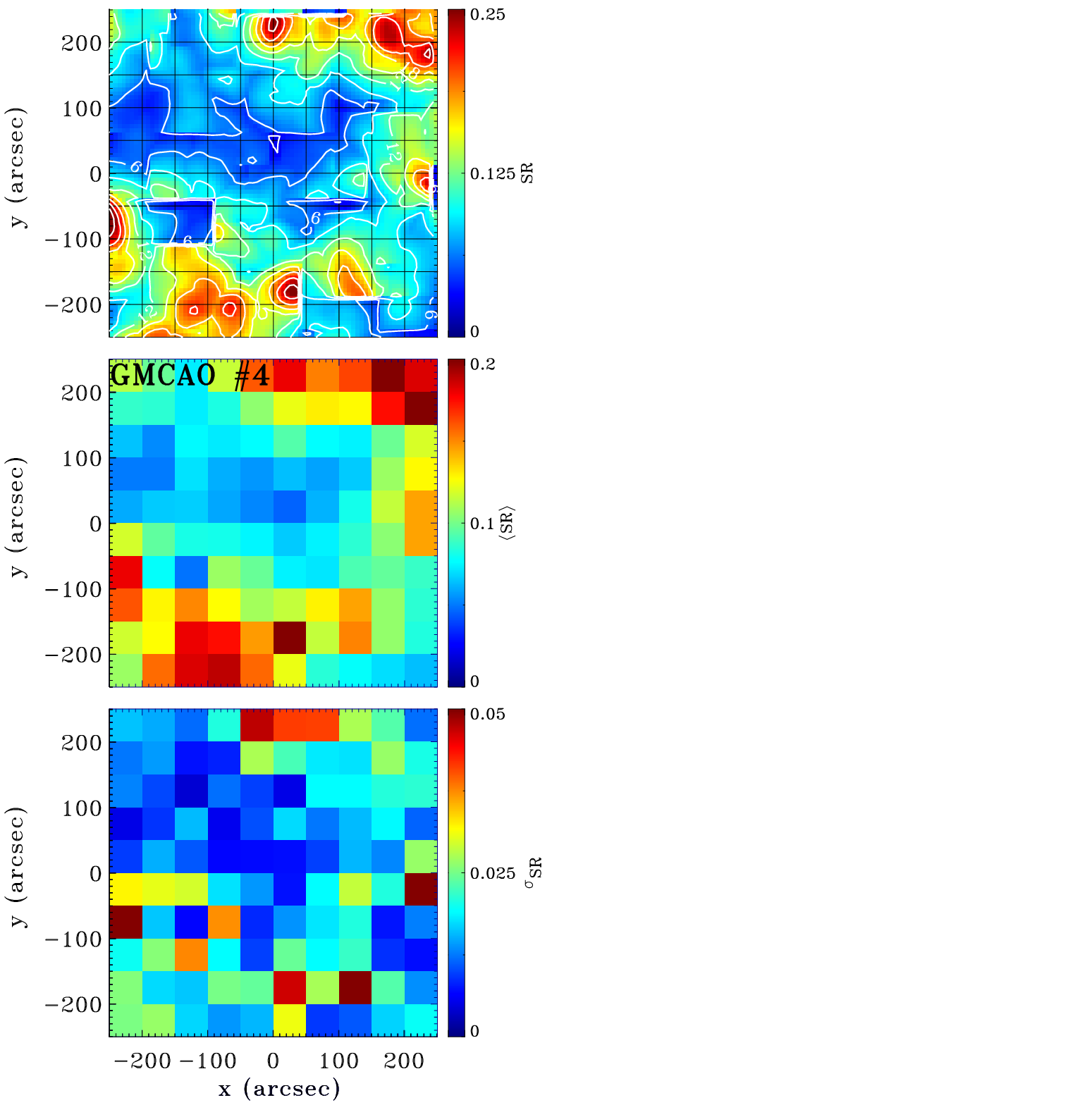}
      \includegraphics[bb=54 355 338 925,width=8.7cm]{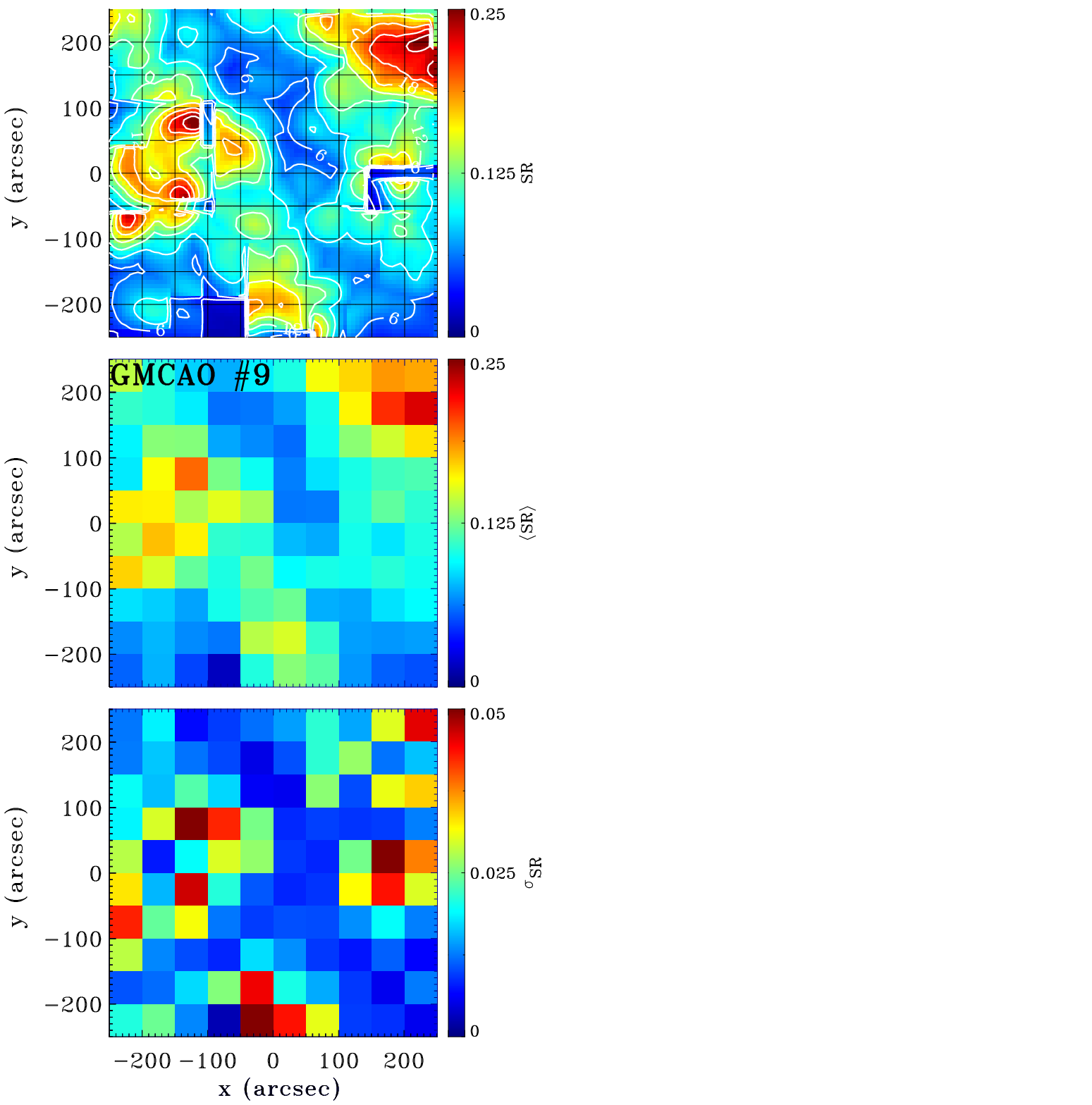}}
\caption{Top panel: $K$-band SR map obtained with GIUSTO. The image is divided into 100 sectors of 50 $\times$ 50 arcsec$^2$ each. Some intensity contours are shown. As discussed in the text, the discontinuities inside the sectors are due to the adopted  dithering technique: sometimes adjacent regions could have rich or poor asterism whenever one of the brighter and closer stars to the scientific FoV star is rejected by our assumptions or lost because of the distance.
  Middle panel: $K$-band map of average SR values measured in each sector. Bottom panel: $K$-band map of the standard deviation of each sector. Color-bars indicate the scales of each panel.}
\label{fig:SR}%
\end{figure*}
\begin{figure*}
\addtocounter{figure}{-1}
\hbox{\includegraphics[bb=54 355 338 925,width=9cm]{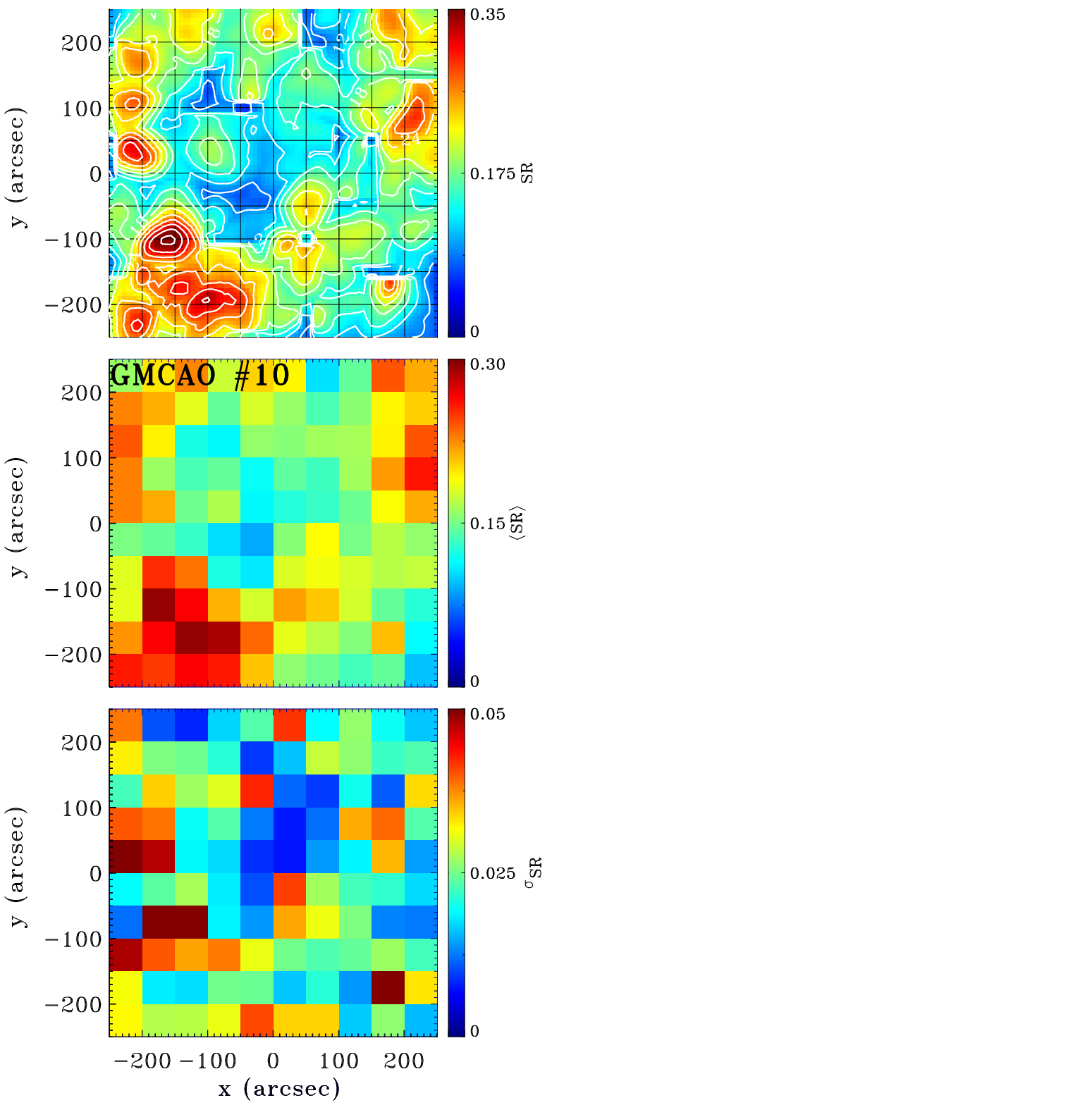}
      \includegraphics[bb=54 355 338 925,width=9cm]{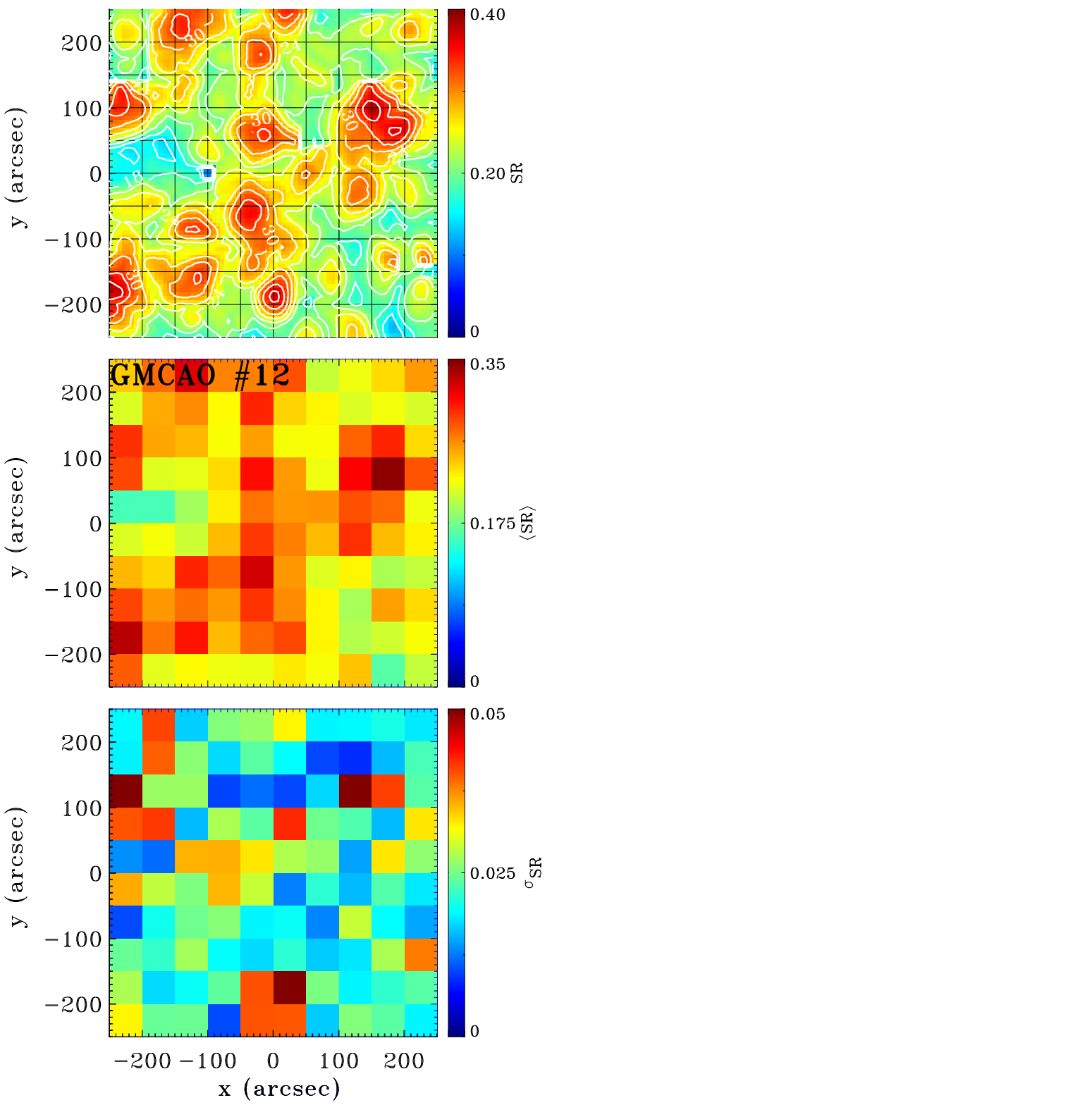}}
\caption{Continued.}
\end{figure*}

\subsection{Discussion}
\label{sec:discussion}
Figure~\ref{fig:histoSR} reports all the average SRs in the various sectors,  i.e., the values referring to the middle panel of Figure~\ref{fig:SR}. The mean is 0.168, bigger (therefore better) than the one found in P17. The maximum value is 0.36, while the standard deviation has a mean of 0.02 and a maximum value of 0.09.

To test the importance of the choice of the used atmospheric profile, we simulated of the GMCAO~\#1 field with the 3 different vertical profiles used to represent the Paranal atmosphere, and Figure~\ref{fig:3chandra} shows the results. The SRs obtained with a 9-layers profile is about 20\% higher than the others, while the differences are less evident between SR retrieved with the 35- and 40- layers atmosphere.

Figure~\ref{fig:histoNGS} shows two histograms of the NGS properties: the $R$-band magnitudes and the radial distance from the center of all sectors of the 22 GMCAO simulations (off-axis). As discussed in Section~\ref{sec:input}, the limit magnitude is $R$=18, and $89\%$ of the stars used as NGSs are fainter than $R=14$ mag, meaning that the GMCAO performance relies on the faint-end part of the stellar distribution.
 We have also superimposed the predicted distribution of stars in the range $8<R<18$ obtained with TRILEGAL \citep{Girardi2005} on those selected in the USNO-B1 catalog for the correction. This population synthesis code that simulates the stellar photometry of our Galaxy is a well-exploited software in the AO community for the investigation and the prediction of the sky coverage. The two distributions are in good agreement.

 The mean off-axis values shown in the histogram, i.e., the average between the off-axis distance of the NGSs, point out that the wide GMCAO Technical FoV is a crucial ingredient for the method. We recall that stars inside the Scientific FoV were rejected, as discussed in  Section~\ref{sec:input}. This plot can give a hint on how wide the Technical FoV should be, as a compromise between performance and feasibility.

Here the FoV limit for the ESO MCAO demonstrator (MAD,\citet{Marchetti2005}) is 120 arcsec, and represented by a dashed line: averaging over the 22 simulations, a MCAO system with such as FoV could access to the 57\% of the stars usable by the larger GMCAO Technical FoV.

The NGS characteristics can be related to the obtained SR, as shown in Figure~\ref{fig:SRax}: while there is no clear correlation with the NGSs mean $R$-band magnitude, there is a trend considering their mean off-axis position. The performance is dominated by how close they are to the Scientific FoV (and how homogeneously the asterism is distributed). We may interpolate the data by a second degree polynomial function ($A_0+A_1x+A_2x^2$) with the following coefficients: $A_0=210.3 \pm2.8$,  $A_1=-585.6\pm34.3$,  $A_2=719.0\pm99.7$.

Figure~\ref{fig:SRstdmean} shows the relation between the SR standard deviation normalized to the mean ($\sigma_{\rm SR}/{<\rm SR>}$) and the mean itself: regions with higher SR tend to have lower standard deviations if compared with the average value.  The relative errors are small for high SR, about 30\%, and rise up to 45\% for the other cases, except for the worst case (field \#1). However, this spread is not so significant, and comparable to those obtained in the fields of P17, and may lead to concluding that the SR distribution is almost uniform. They are also in good agreement with the results of \citet{Arcidiacono2003} presenting the study of the SR uniformity in a Layer-Oriented MCAO.

\begin{figure*}
  \center
\includegraphics[bb=59 790 759 1175,width=17cm]{figs/surveyscomp.eps}
\caption{Histograms of average (left-hand panel) and standard deviation (right-hand panel) of SR in all sectors of the 22 GMCAO fields. Dashed lines represent the mean values.}
\label{fig:histoSR}
\end{figure*}

\begin{figure*}
  \center
\includegraphics[bb=109 395 729 1140,width=7cm]{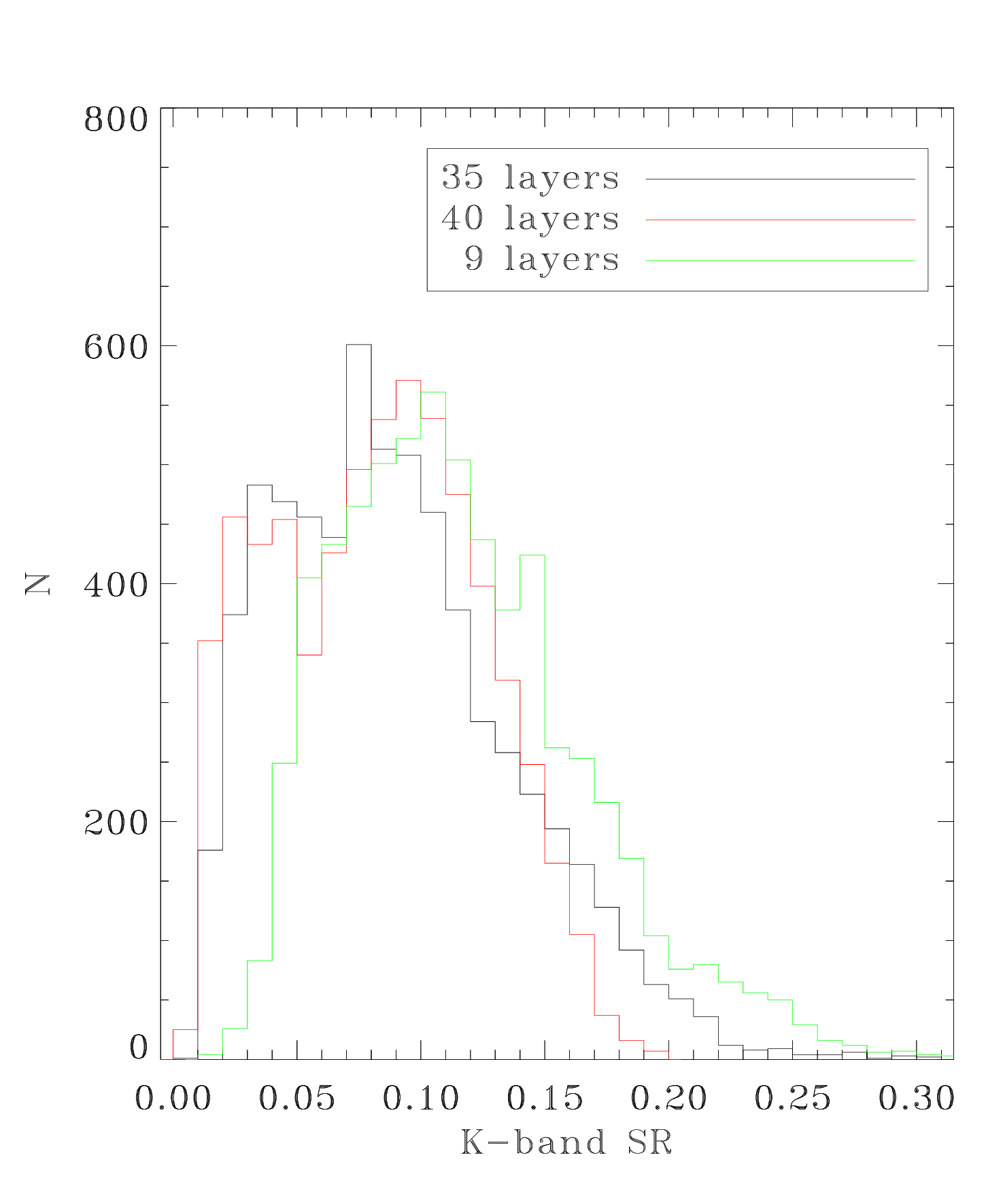}
\caption{ Histograms of average SR in all sectors of the GMCAO\#1 field for 3 different atmosphere profiles: the one used in this work (35 layers, black line), in the previous P17 (40 layers, red line), and the one adopted in some other works (9 layers, green line).}
\label{fig:3chandra}
\end{figure*}

\begin{figure*}
\center
\includegraphics[bb=50 790 759 1175,width=17cm]{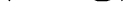}
\caption{Histograms of the NGS $R$-band magnitudes (left-hand panel) and radial distance from the central pointing (right-hand panel) for all the 100 sectors investigated of the 22 GMCAO simulations. The red distribution represents the values we found using TRILEGAL. The dotted line represents the lower limit for finding NGSs in the technical field, as explained in Section~\ref{sec:input}. The dashed line represents the limit of the Technical FoV for a 'usual' MCAO system.}
\label{fig:histoNGS}
\end{figure*}

\begin{figure*}
\center
\includegraphics[bb=57 787 759 1180,width=17cm]{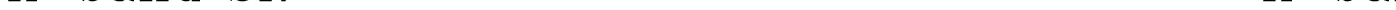}
\caption{NGS $R$-band magnitude (left-hand panel) and radial distance from the central pointing (right-hand panel) as a function of the SR of the field for all the 22 GMCAO fields. The solid line represents the best-fit, while the dashed one shows the boundary of the Technical field. }
\label{fig:SRax}
\end{figure*}

We considered the median value of SR$=0.127$, derived in P17, as the threshold to discern whether a sector in the $10\times 10$ grids is successfully observable or not.
We also define the threshold SR=0.08 as the limit value useful to detect late-type galaxies with $M\le10^9$ M$_{\odot}$ at a redshift $z\ge2.75$, as discussed in Section~\ref{sec:intro}. Therefore the sectors with SRs less than that value can be tagged as unsuccessful for source detection. Figure~\ref{fig:SRsuccess}, based on results shown in Table~\ref{tab:SRresults} shows that 11 fields out of 22 (half of the sample) have a percentage of success between 80\% and 100\%. Moreover, 11 fields show a percentage of un-success between 0 and 5\%. These are the most promising candidates for a complete study, as done in P17, i.e., for source detection and photometric decomposition.

\begin{figure}
\center
\includegraphics[bb=60 380 759 1210,width=7cm]{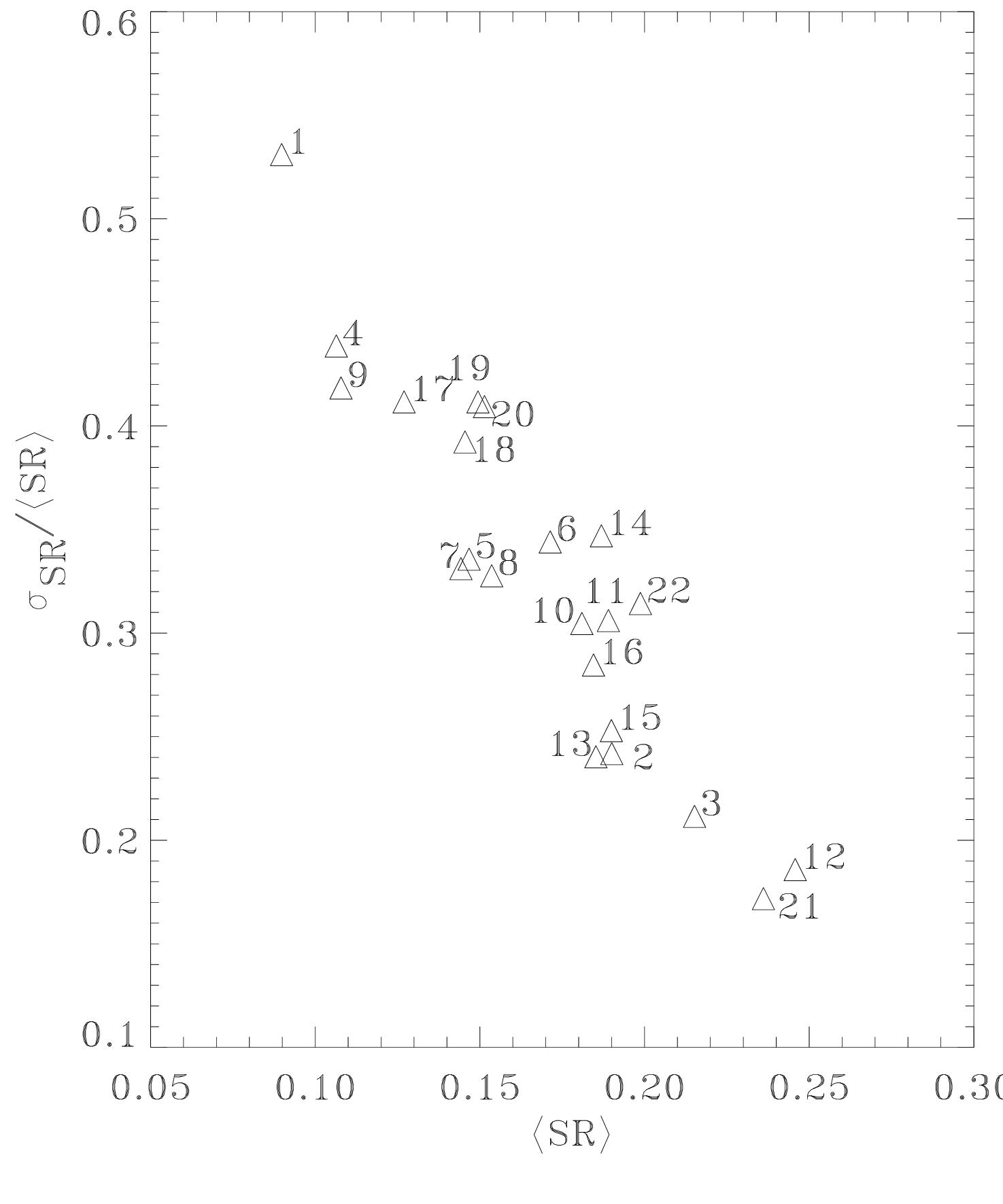}
\caption{SR standard deviation, normalized to the corresponding average SR value, as a function of the average SR itself.}
\label{fig:SRstdmean}
\end{figure}

\begin{figure}
  \center
\includegraphics[bb=77 397 750 1210,width=7cm]{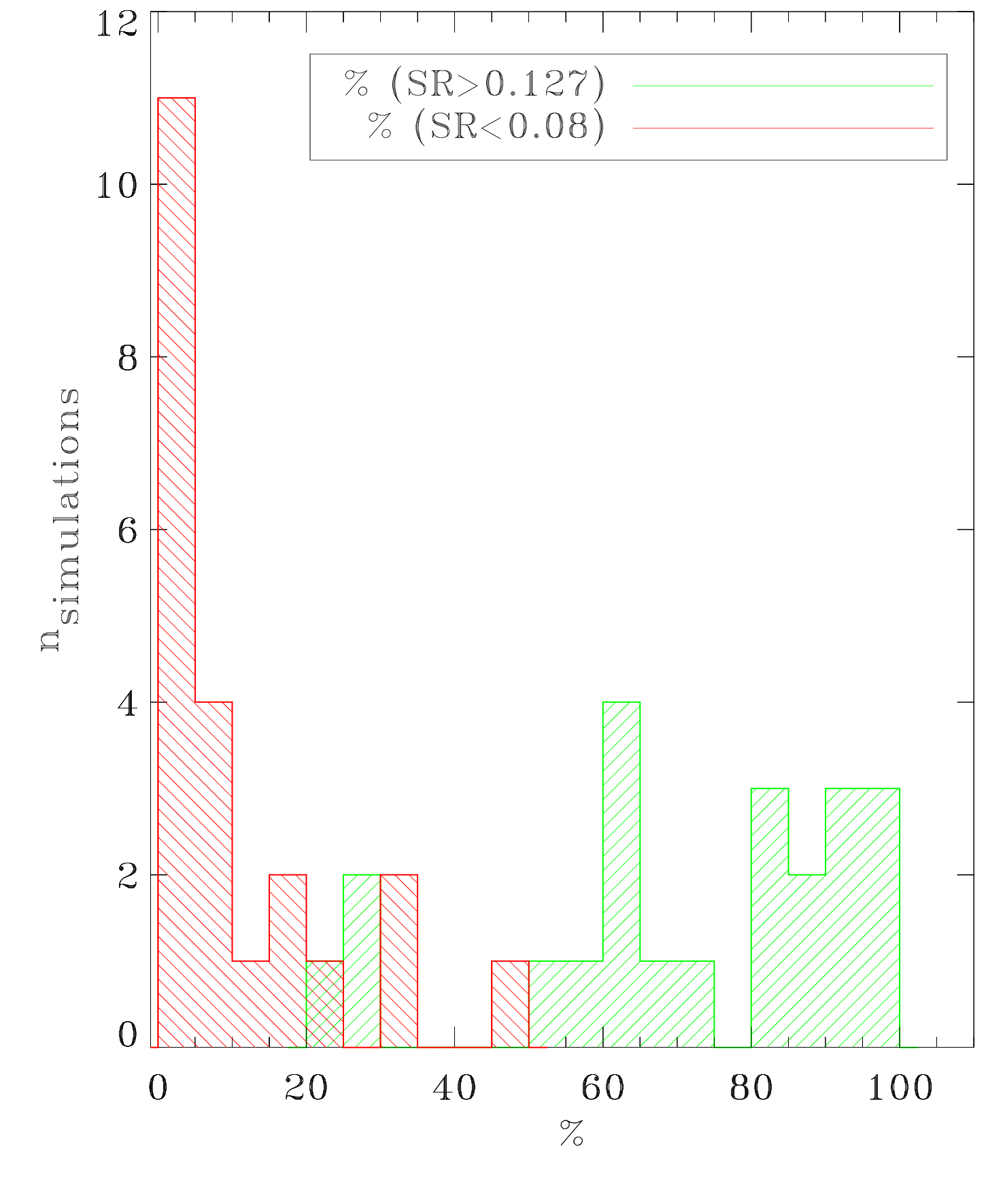}
\caption{Green: histogram of the percentage of the successes, i.e., for each region how many SR values are greater than 0.127, as discussed in the text. Red: histogram of the percentage of the un-success, i.e., how many SR values are less than 0.08, as discussed in the text.} 
\label{fig:SRsuccess}
\end{figure}

\begin{figure}
  \center
\includegraphics[bb=37 357 750 1210,width=8.5cm]{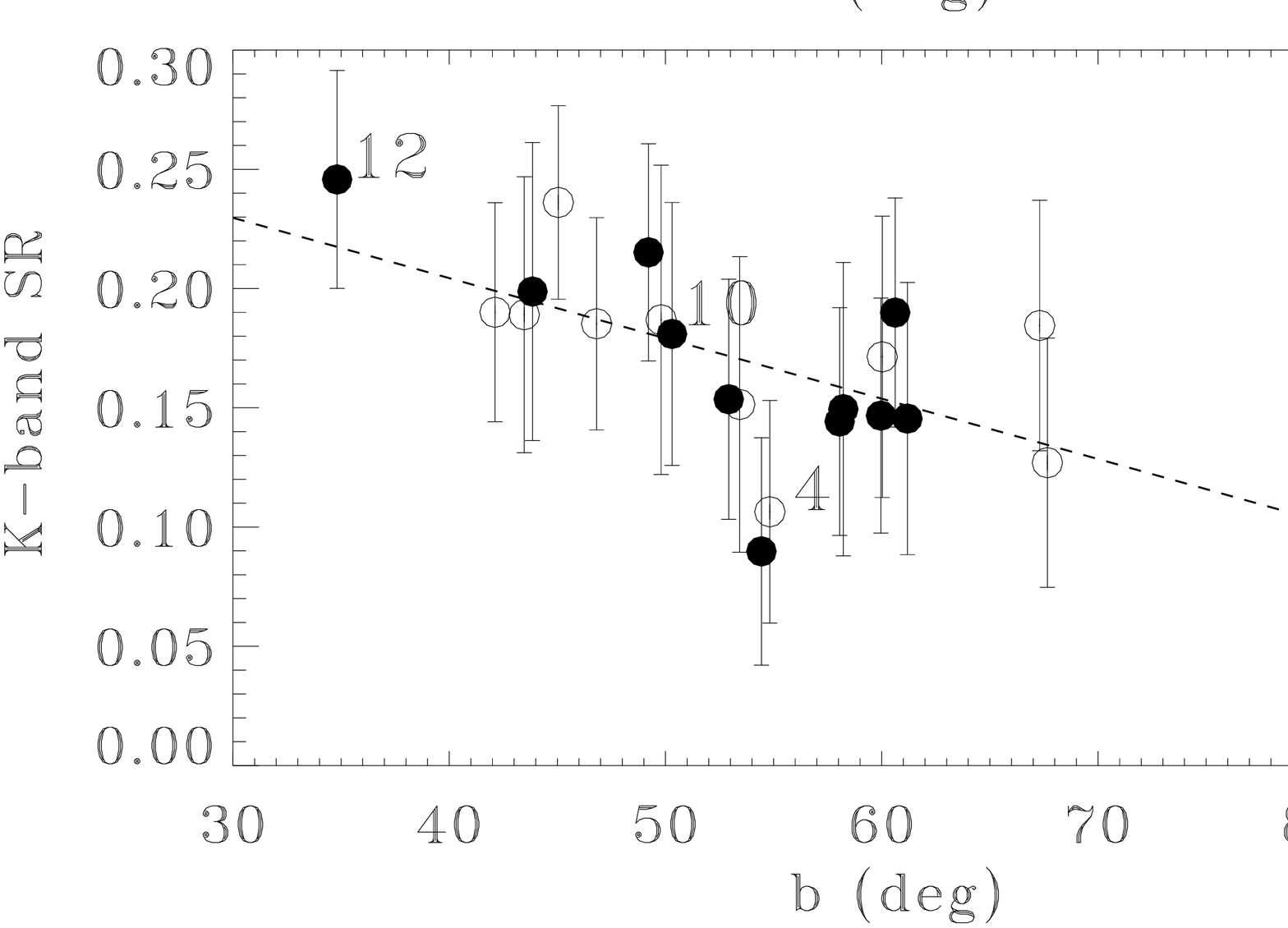}
\caption{$K$-band SR as a function of the galactic coordinates of the 22 GMCAO fields: longitude (top panel) and latitude (bottom panel). Filled circles represent points with negative values of galactic latitude, while open circles are for positive values. The error bars represent the one $\sigma$ value.}
\label{fig:SRcoo}
\end{figure}

When we consider the galactic coordinates of the 22 simulations, we find a correlation between the $K$-band SR and the latitude of the fields, as shown in Figure~\ref{fig:SRcoo}. A linear polynomial function with coefficients $B_0=30.55\pm3.59$ and $B_1=-0.25\pm0.06$ can fit this relation. Few fields exceed the expected value of this relation, one of them is the GMCAO~\#12 (DLS-3 survey), where almost all the sectors have ${\rm SR}>0.127$. On the contrary, poor results are obtained in GMCAO~\#4 (HDF survey), together with the GMCAO~\#1 (CDFS survey, already presented in P17), where less than 30\% of the region has ${\rm SR}>0.127$ and more than 30\% has ${SR}<0.08$. GMCAO~\#10 (DLS-1 survey) has a ``median'' behavior, which perfectly follows the relation. We also highlighted GMCAO~\#9 field (SubaruDF survey) because it is the northernmost region, being at $\approx 83^{\circ}$. Therefor the sub-sample of 4 simulations shown in Figure~\ref{fig:SR} (GMCAO \#4, \#9, \#10, \#12) can be considered as representative of different cases of this work.

\section{Conclusions}
\label{sec:conclusion}
 In this work, we are testing an innovative AO method, called GMCAO, which takes advantage of a wide Technical FoV for finding NGSs. It can be used for a 40-m new generation telescope as an alternative, or a back-up solution, to LGS-based AO systems for deep-field observations. In this paper we carryed out simulations of 22 fields, increasing the number of target fields with respect to the single case presented in EP17, improving the statistics in terms of performance estimation. Our results can be summarised as follows.
 \begin{enumerate}
     \item We have selected 22 regions in the sky from a list of the most studied deep-field surveys, and we have investigated the performance of the GMCAO approach in terms of sky coverage.  The selection did not include observational criteria, in the sense that we assumed that an ELT-like instrument  could observe all the fields at the same atmospherical and operational condition, i.e., at a zenithal distance of $\theta =30\deg$. In this way, we avoid introducings a bias in our results using different values of the Fried parameter $r_0$, as it varies with the $[cos(\theta)]^{3/5}$ \citep{Beckers1993}.
      \item  We used GIUSTO, an IDL tomographic simulation tool that, once defined several technical and astrophysical input parameters, calculates the performance of a given telescope that benefits from GMCAO in terms of SR over a given scientific FoV. We ran GIUSTO on $10\times 10$ sectors of $50\times50$ arcsec$^2$ each to match the MICADO FoV and for all the 22 fields,  for a total of about 1528 arcmin$^2$. 
     \item We configured GIUSTO to optimize the VDMs conjugation altitudes, which depend on the guide stars constellation geometry. We derived the statistically optimal altitudes to which conjugate the 6 VDMs, given the 35-layers profile adopted:  0, 3.7, 9.1, 11.7, 15.9, and 24.4~km, respectively. 
     \item We built $500\times500$ arcsec$^2$ SR maps for all the simulations, measuring the average SR for each sector and the corresponding standard deviation.
        The overall average SR value of the entire work is $0.168\pm0.023$, which is higher than the one in our previous work, $0.127\pm0.050$.
 We recalls that the atmosphere profiles of the two works are different: here, we used 35 layers, which became the ESO-suggested Paranal profile, while P17 used 40 layers, which was the former baseline for adaptive optics numerical simulation of the ESO ELT. However, the differences between SRs obtained using those two assumptions would not play a key role.
 Considering other findings of P17, we have also investigated here the percentage of success and un-success of the GMCAO simulations using the results of our previous work. The 72$\pm23\%$ of the overall sky covered by the GMCAO fields has ${\rm SR}$ larger than limit of $0.127$ measured in EP17 for successful photometric and morphological analysis of galaxies at high redshifts, and only $\pm13\%$ has ${\rm SR}<0.08$.
 
\item We noted that the SR values found anti-correlate linearly with the galactic latitude. This relation was expected because of the distribution of star density scales with the galactic latitudes.
We found, predictably, that the CDFS is the most challenging region for AO observations. The SubaruDF is the closest to the North Galactic Pole and is the second in terms of difficulty.
The best performance is obtained using stars that are relatively close to the Scientific FoV, suggesting a correlation between the SR and the mean off-axis position of NGSs. The magnitude does not seem to be a relevant parameter: we can speculate that the number of the stars (or equivalently the sky coverage) is high enough to allow optimal star selection. 
We demonstrated that GMCAO might achieve larger sky coverage than, i.e., the MAD system using NGS solely that worked with the MCAO approach.
\end{enumerate}
The outcome of our study confirms the performance achievable by the GMCAO technique allowing both source detection and photometric decomposition. Therefore, GMCAO can help to characterize objects at high redshifts and add a piece of information in the puzzle of the formation and evolution of galaxies. In this sense, GMCAO represents a robust alternative way or a risk mitigation strategy to a MCAO laser-based system for the new generation of ELTs.

\section{Acknowledgments}
We acknwledge the anonymous Referee that helped to improve the manuscript and contributed, among others, to make more accurate the discussion about the WFS error scaling.
EP has been supported by the Italian Ministry for Education University and Research (MIUR) under the grant Premiale ADONI 2016.

\bibliographystyle{aas}

\end{document}